\documentclass[aps,prl,twocolumn,showpacs,amsmath,amssymb,superscriptaddress]{revtex4-1}

\usepackage{graphicx}
\usepackage{dcolumn}
\usepackage{bm}
\usepackage{color}
\usepackage{graphicx}
\usepackage{epsfig}
\usepackage{fancyhdr}
\usepackage{epstopdf}
\include{graphic}

\begin{document}


\title{Coherent Microwave Emission of a Gain-Driven Polariton}


\author{Bimu~Yao}
\affiliation{Department of Physics and Astronomy, University of Manitoba, Winnipeg, Canada R3T 2N2}
\affiliation{State Key Laboratory of Infrared Physics, Chinese Academy of Sciences, Shanghai 200083, People's Republic of China}
\affiliation{School of Physical Science and Technology, ShanghaiTech University, Shanghai 201210, China}

\author{Y.S.~Gui}
\affiliation{Department of Physics and Astronomy, University of Manitoba, Winnipeg, Canada R3T 2N2}

\author{J.W.~Rao}
\affiliation{Department of Physics and Astronomy, University of Manitoba, Winnipeg, Canada R3T 2N2}
\affiliation{School of Physical Science and Technology, ShanghaiTech University, Shanghai 201210, China}

\author{Y.H.~Zhang}
\affiliation{Department of Physics and Astronomy, University of Manitoba, Winnipeg, Canada R3T 2N2}

\author{Wei~Lu}
\affiliation{State Key Laboratory of Infrared Physics, Chinese Academy of Sciences, Shanghai 200083, People's Republic of China}
\affiliation{School of Physical Science and Technology, ShanghaiTech University, Shanghai 201210, China}

\author{C.-M.~Hu} \email{hu@physics.umanitoba.ca; \\URL: http://www.physics.umanitoba.ca/$\sim$hu}
\affiliation{Department of Physics and Astronomy, University of Manitoba, Winnipeg, Canada R3T 2N2}

\date{\today}

\begin{abstract}
By developing a gain-embedded cavity magnonics platform, we create gain-driven polariton (GDP) that is activated by an amplified electromagnetic field. Distinct effects of gain-driven light-matter interaction, such as polariton auto-oscillations, polariton phase singularity, self-selection of a polariton bright mode, and gain-induced magnon-photon synchronization, are theoretically studied and experimentally manifested. Utilizing the gain-sustained photon coherence of the GDP, we demonstrate polariton-based coherent microwave amplification ($\sim$ 40 dB) and achieve high-quality coherent microwave emission (Q $>$ 10$^9$).

\end{abstract}

\maketitle

One of the recent spotlights in the realm of light-matter interaction is the renaissance of the cavity magnon polariton (CMP) \cite{Artman1953,Soykal2010,Huebl2013,Tabuchi2014,Zhang2014,Bai2015}. It has led to the emergence of cavity magnonics \cite{Rameshti2021} that uses the magnon as a key component for developing classical and quantum computing systems \cite{Lachance-Quirion2019}. Over the past decade, CMP has attracted broad interest from communities studying quantum information \cite{Zhang2019}, cavity QED \cite{Clerk2020}, and cavity optomechanics \cite{Zhang2016,Osada2016,Potts2021}. It has enabled the innovation of magnonic memories \cite{Zhang2015a,Shen2021}, cavity spintronics \cite{Bai2015,Bai2017}, quantum sensors \cite{Lachance-Quirion2020}, and a variety of transducers \cite{Haigh2016,Lachance-Quirion2019}. Intriguing effects such as polariton bistability \cite{Wang2018}, level attraction \cite{Harder2018,Bhoi2019,Boventer2019}, exceptional points \cite{Zhang2017}, Floquet ultrastrong coupling \cite{Yao2017,Xu2020}, unidirectional invisibility \cite{Wang2019}, and bound states in the continuum \cite{Yang2020} have been observed in cavity magnonics. All these progresses stem from the hybrid nature of the CMP, which enables using photons to control magnons and vice versa. However, this benefit comes at a cost: the light-matter interaction, in general, produces non-degenerate polariton states, so that photon states depend on the superposition of polaritons oscillating at different frequencies with different phases and damping rates. This effect causes amplitude modulation and phase aberration for photons, hindering photonic applications that require stable monochromatic operation with well-sustained photon coherence.

Can light-matter hybridization, monochromatic operation, and sustained photon coherence be engineered to coexist in polaritonics? The answer is yes, but the only example that nature has taught us is Bose-Einstein condensation (BEC) of cavity exciton polariton \cite{Deng2002,Kasprzak2006,Deng2010}. Although polariton BEC has invigorated polaritonics \cite{Deng2010} and transformed laser technology \cite{Schneider2013}, it is not a silver bullet since BEC is not easy to realize in most polariton systems. Another approach, proposed very recently, was to integrate magnon-photon coupling with the spin-torque effect, generating macroscopic phase coherence in spin-torque oscillators \cite{Hou2021}. Yet, this intriguing idea awaits experimental validation.

In this Letter, by developing gain-embedded cavity magnonics, we reveal a new path for achieving monochromatic polariton operation with sustained photon coherence. A gain-driven polariton (GDP) is created, which exhibits distinct features such as polariton phase singularity, polariton auto-oscillations, and spontaneous polariton mode selection. Using these properties, we demonstrate polariton-based coherent microwave amplification and high-quality microwave emission.

\begin{figure} [ht!]
\begin{center}
\includegraphics[width=\columnwidth]{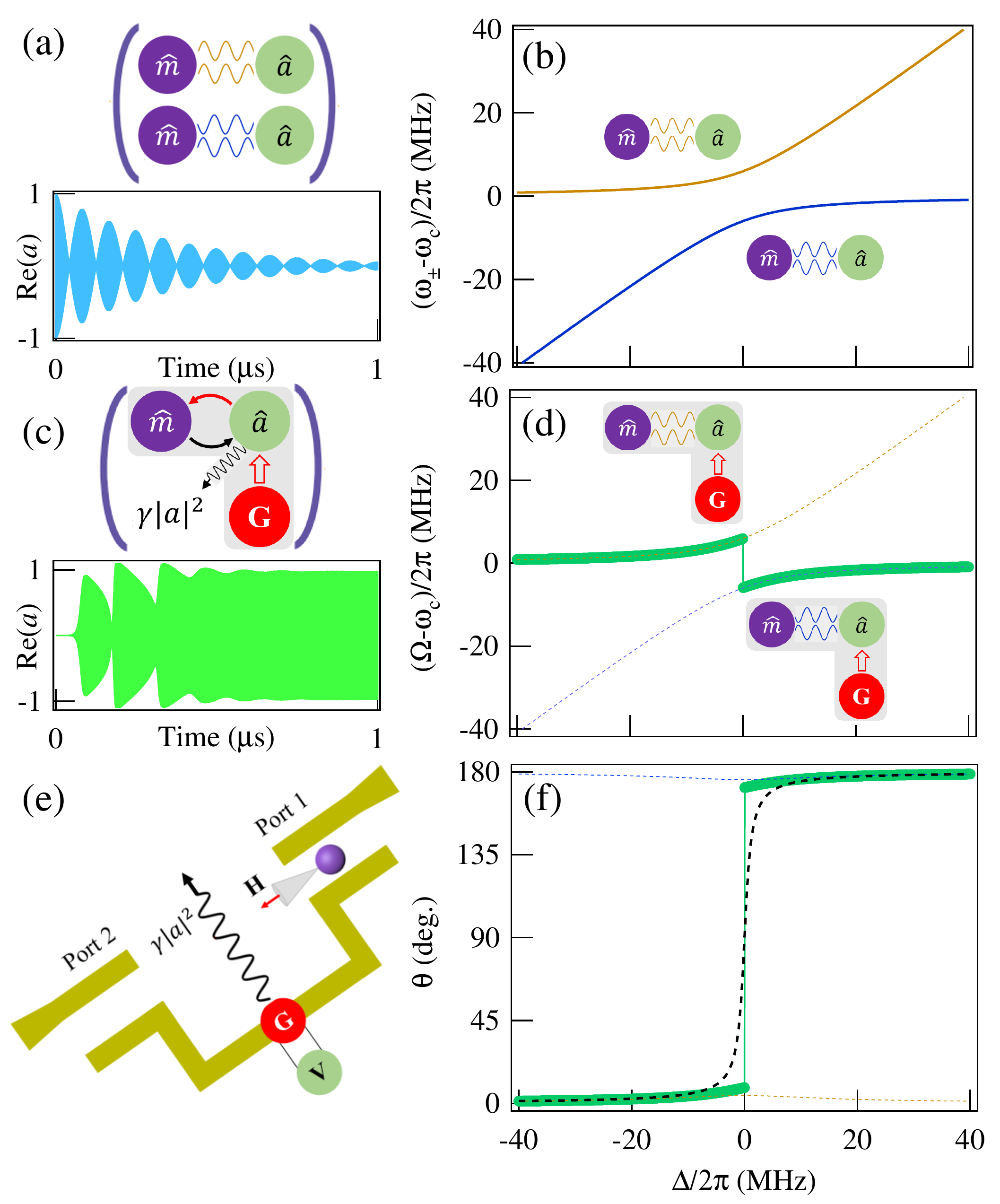}
\caption{\textbf{CMP vs GDP.} (a) Conventional CMP modes characterized by Rabi oscillations and (b) level repulsion. (c) GDP that leads to a photon steady-state and (d) self-selection of a bright polariton mode. (e) Schematic device architecture showing the one-photon gain $G$ and the two-photon damping rate $\gamma |a|^2$. (f) For closed cavity magnonics, the phase of the GDP (green dots) exhibits a singularity at $\Delta$ = 0, resembling the phase of a driven oscillator (black dashed curve). The phases of the two CMP modes (orange and blue dotted curves) are plotted for comparison.}\label{Fig1}
\end{center}
\end{figure}

To explain the idea of the GDP, let us start from the CMP produced in closed cavity magnonics governed by the coherent magnon-photon coupling \cite{Rameshti2021}:
\begin{eqnarray}
\dot{\hat{a}}&=&-i\widetilde{\omega}_c\hat{a}-iJ\hat{m}, \nonumber \\
\dot{\hat{m}}&=&-i\widetilde{\omega}_m\hat{m}-iJ\hat{a}.
\label{e1}
\end{eqnarray}
Here, $\hat{a}$ and $\hat{m}$ represent the annihilation operator of the cavity photon and magnon modes, respectively. $J$ is the rate of the coherent coupling. $\widetilde{\omega}_c=\omega_c-i\kappa_c$ and $\widetilde{\omega}_m=\omega_m-i\kappa_m$ correspond to the complex frequencies of the cavity and magnon modes, respectively, where $\kappa_c$ and $\kappa_m$ are the damping rates. The solution of Eq.~\ref{e1} is a pair of CMP modes $\hat{p}_{\pm} = c_{\pm}\hat{a} \pm c_{\mp}\hat{m}$ with eigenfrequencies $\widetilde{\omega
}_{\pm} = \widetilde{\omega}_c+\widetilde{\Delta}/2 \pm \widetilde{\Lambda}$, where $c_{\pm} =[(\widetilde{\Lambda} \mp \widetilde{\Delta}/2)/2\widetilde{\Lambda}]^{1/2}$ are the state amplitudes with $\widetilde{\Delta} \equiv \widetilde{\omega}_m - \widetilde{\omega}_c$ and $\widetilde{\Lambda} \equiv [J^2 + (\widetilde{\Delta}/2)^2]^{1/2}$. They show in-phase and out-of-phase oscillations of the magnons and photons as depicted in Fig.~\ref{Fig1}(a) and (b).

The GDP is created by incorporating two new ingredients (gain and nonlinearity) into the cavity, changing its complex frequency to
\begin{equation}
\widetilde{\omega}_c=\omega_c+i(G-\kappa_c-\gamma |a|^2).
\label{e2}
\end{equation}
Here, $G$ represents the one-photon gain, $a$ is the expectation value of $\hat{a}$, and $\gamma |a|^2$ is the two-photon (nonlinear) damping rate. Setting $G > \kappa_c$ and switching on the gain, $|a(t)|$ exponentially increases to the steady state value $|A_c|$, where the nonlinear damping balances the net gain with $G-\kappa_c = \gamma |A_c|^2$. At the steady state, the cavity becomes an auto-oscillator with zero-damping populated with a constant number of photons $n_c = |A_c|^2$. Such a cavity resembles the gedanken oscillator conceived in 1920 by van der Pol \cite{VdP1920}, which has been broadly used for understanding nonlinear biological processes and quantum dynamics \cite{Lee2013,Dutta2019}.

As sketched in Fig.~\ref{Fig1}(c), when coupling the zero-damping cavity with a damped oscillator, such as the magnon mode, an intriguing gain-driven light-matter interaction takes place. Mathematically, Eqs.~\ref{e1} \& \ref{e2} have a steady-state solution \cite{supp} with the expectation values $a(t) = Ae^{-i\Omega t}$ and $m(t) = M e^{-i\Omega t} e^{-i\theta}$, where $A$ and $M$ are the steady state amplitude of photons and magnons, respectively. At the steady state, magnons and photons are synchronized at the frequency $\Omega$ with a relative phase $\theta$ that depends on the detuning $\Delta = \omega_m - \omega_c$. Meanwhile, the CMP states $\hat{p}_{\pm} = c_{\pm}\hat{a} \pm c_{\mp}\hat{m}$ evolve to the GDP with the steady state amplitudes
\begin{equation}
P_{\pm} = c_{\pm}A \pm c_{\mp} e^{-i\theta}M.
\label{e3}
\end{equation}

\begin{figure} [t!]
\begin{center}
\includegraphics[width=\columnwidth]{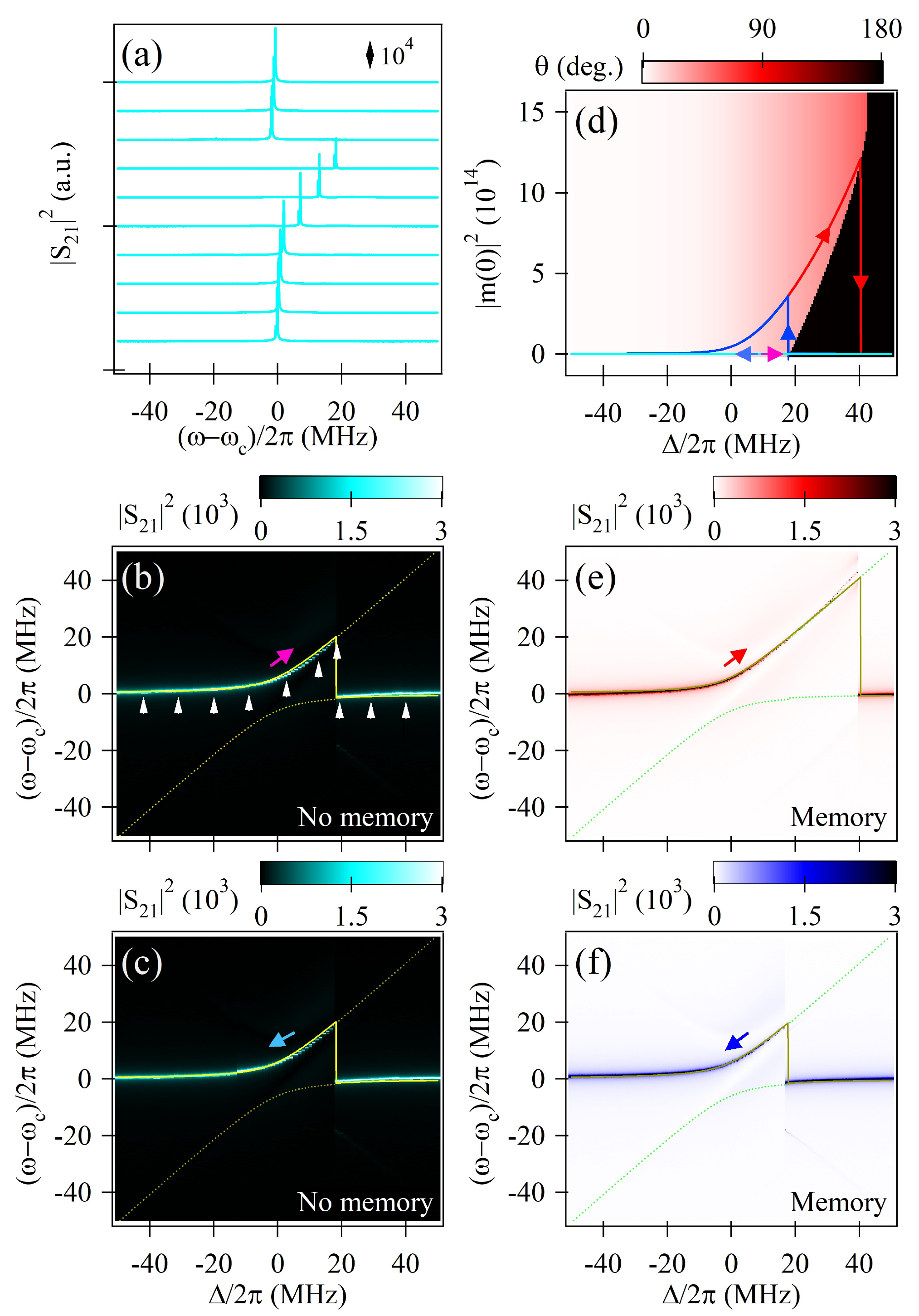}
\caption{\textbf{Phase singularity.}  (a) $|S_{21}(\omega)|$ measured in the \textquotedblleft no-memory" setting by sweeping $\omega$, with the detuning increased from -42 to 40 MHz. The dispersion of the bright GDP mode, measured by (b) increasing and (c) decreasing $\Delta$ in the \textquotedblleft no-memory" setting. The vertical arrows in (b) mark the detunings for the spectra presented in (a). (d) Numerically calculated $\theta(\Delta)$ at different initial conditions, where the sharp boundary marks the phase singularity. Solid curves correspond to $|m(0)|^2$ calculated for different sweeping settings. (e) The dispersion measured by increasing and (f) decreasing $\Delta$ in the \textquotedblleft memory" setting. The solid and dotted curves in (b),(c) and (e),(f) are the fitted GDP and CMP dispersions, respectively.}\label{Fig2}
\end{center}
\end{figure}

Physically, due to the nonlinear term in Eq.~\ref{e2}, the phase $\theta$ is set by the initial condition for Eq.~\ref{e1}. Solving Eqs.~\ref{e1} \& \ref{e2} numerically \cite{supp}, we find two cardinal features of the GDP: (1) a phase singularity emerges at the critical detuning $\Delta_c$, where $\theta$ switches abruptly from $\theta \approx 0$ to $\theta \approx \pi$. In general, as we will see later in Fig. \ref{Fig2}, $\Delta_c$ depends on the initial condition and the coupling parameters, but in the simplest case for the closed cavity magnonics operating with the initial condition $a(0) \sim 0$ and $m(0) \sim 0$, we find $\Delta_c = 0$, where the calculated $\theta(\Delta)$ plotted in Fig.~\ref{Fig1}(f) indicates that the GDP has both characteristics of a driven oscillator and polariton. (2) Along with the phase singularity, the GDP spontaneously selects a bright mode \cite{supp}:
\begin{subequations}
\begin{align}
|P_-| = 0,~ |P_+| > 0,~\Omega \approx \omega_+,~~ for~ \Delta < \Delta_c;  \\
|P_+| = 0,~ |P_-| > 0,~\Omega \approx \omega_-,~~ for~ \Delta > \Delta_c.
\end{align}
\label{e4}
\end{subequations}
Equation~\ref{e4} means the GDP is composed of a bright and a dark polariton mode, whose frequencies abruptly interchange at $\Delta_c$ as shown in Fig.~\ref{Fig1}(d). By spontaneously selecting the bright mode, the magnon and photon are in-phase synchronized at $\omega_+$ for $\Delta < \Delta_c$, but they become out-of-phase synchronized at $\omega_-$ for $\Delta > \Delta_c$. Hence, at any detuning, the GDP unifies light-matter hybridization with monochromatic auto-oscillation. This observation leads to sustained photon coherence, opening new avenues for polariton applications, as we will demonstrate later.

Experimentally, we design gain-embedded cavity magnonics to realize our idea. Figure~\ref{Fig1}(e) shows the schematic picture of our device: A half-wavelength microstrip line resonator is connected with two measurements ports, and a voltage-controlled amplifier is engineered on the microstrip to amplify the cavity mode. A yttrium iron garnet (YIG) sphere with a diameter of 1~mm is mounted on a probe attached to an x-y-z stage so that its position is controlled with the resolution of 5 $\mu$m.  Setting the YIG sphere on the cavity and applying a magnetic field $\mathbf{H}$ to control the Kittel mode frequency $\omega_m$=$\gamma_e\mu_0(|H|+H_A)$, we measure the microwave transmission and emission spectra at polariton steady states. Here, $\gamma_e$ = $2\pi\times27.3$~GHz/T is the electron gyromagnetic ratio, $\mu_0 H_A$ = -6.2~mT is the effective field of YIG, and $\mu_0$ is the vacuum permeability. A dozen devices with different geometries are measured by placing the YIG sphere at different positions, all of which exhibit the same feature as summarized by Eq.~\ref{e4}. The data presented below are rendered from one of our typical devices with parameters $\omega_c/2\pi$ = 3.588~GHz, $\kappa_m /2\pi$ = 0.9~MHz, $\kappa_c/2\pi$ = 142~MHz, and $\gamma/2\pi$ = 2.6$\times 10^{-12}$ MHz \cite{supp}. In this gain-embedded device, in addition to the coherent coupling rate of $J /2\pi$ = 4.5~MHz, the insertion of the amplifier exposes magnons and cavity photons to the environmental photon bath, which induces a dissipative magnon-photon coupling \cite{Wang2019} with the rate $\Gamma/2\pi$ = 6.1~MHz. This effect shifts $\Delta_c$ from zero, as was displayed in Fig.~\ref{Fig1}(d) for the simple case of the closed cavity magnonics, but as we will see from the experimental data presented in Fig. \ref{Fig2}, it does not change the cardinal feature summarized by Eq.~\ref{e4}. The gain is controlled by a DC voltage $V$. Unless specified, the data are measured at $V$ = 7 V with $G/2\pi$ = 312~MHz.

Figure \ref{Fig2}(a) shows the microwave transmission spectra $|S_{21}(\omega)|$ measured by using the vector network analyzer to sweep the frequency $\omega$. The power level of the probe signal is kept at -50 dBm. As predicted by Eq.~\ref{e4}, instead of level repulsion of twin CMP, only one bright GDP mode is observed, which suddenly changes its mode frequency at $\Delta_c$. The dispersion of the bright GDP mode, measured by either increasing or decreasing $\Delta$, are plotted in Fig. \ref{Fig2}(b) and (c), respectively. In both cases, we determine $\Delta_c/2\pi$ = 18 MHz.

To quantitatively analyze $\Delta_c$ in our devices, we numerically calculate $\theta(\Delta)$ by generalizing Eq.~\ref{e1} to open cavity magnonics \cite{supp}. The obtained phase mapping, by setting $a(0) = A_c$ and changing $m(0)$, is plotted in Fig. \ref{Fig2}(d), where the sharp boundary marks the phase singularity. The experimental data of Figs. \ref{Fig2}(a) - (c) are measured in the \textquotedblleft no-memory" setting: at each detuning, the gain is switched on before measuring $|S_{21}(\omega)|$, and then it is switched off before changing to the next $\Delta$. This setting corresponds to the initial condition of $m(0) = 0$ at each detuning, as indicated by the horizontal line in Fig. \ref{Fig2}(d), where we find $\Delta_c/2\pi$ = 18 MHz.

Changing the initial condition $m(0)$, $\Delta_c$ shifts as shown in Fig. \ref{Fig2}(d). To verify such nonlinear characteristics, we further measure the microwave transmission in the \textquotedblleft memory" setting, where the gain is kept on while changing $\Delta$. In this setting, as shown by the red and blue curves in  Fig.~\ref{Fig2}(d), $m(0)$, at the next detuning \textquotedblleft remembers" the steady-state value of the previous detuning, so that $\Delta_c$ should depend on the sweeping history. Indeed, as shown in Figs.~\ref{Fig2}(e) and (f), the measured phase singularity appears at $\Delta_c/2\pi$ = 40 and 18 MHz for up and down sweeping, respectively. Using the same set of device parameters \cite{supp}, the fitted GDP dispersions (solid curves) agree very well with the measured dispersions in all cases that are shown in Figs.~\ref{Fig2}(b),(c) and (e),(f). Thus, we have created the GDP that spontaneously selects a bright polariton mode. Such a unique polariton enables gain-sustained photon coherence, as we demonstrate below in photon emission experiments.

\begin{figure} [t!]
\begin{center}
\includegraphics[width=\columnwidth]{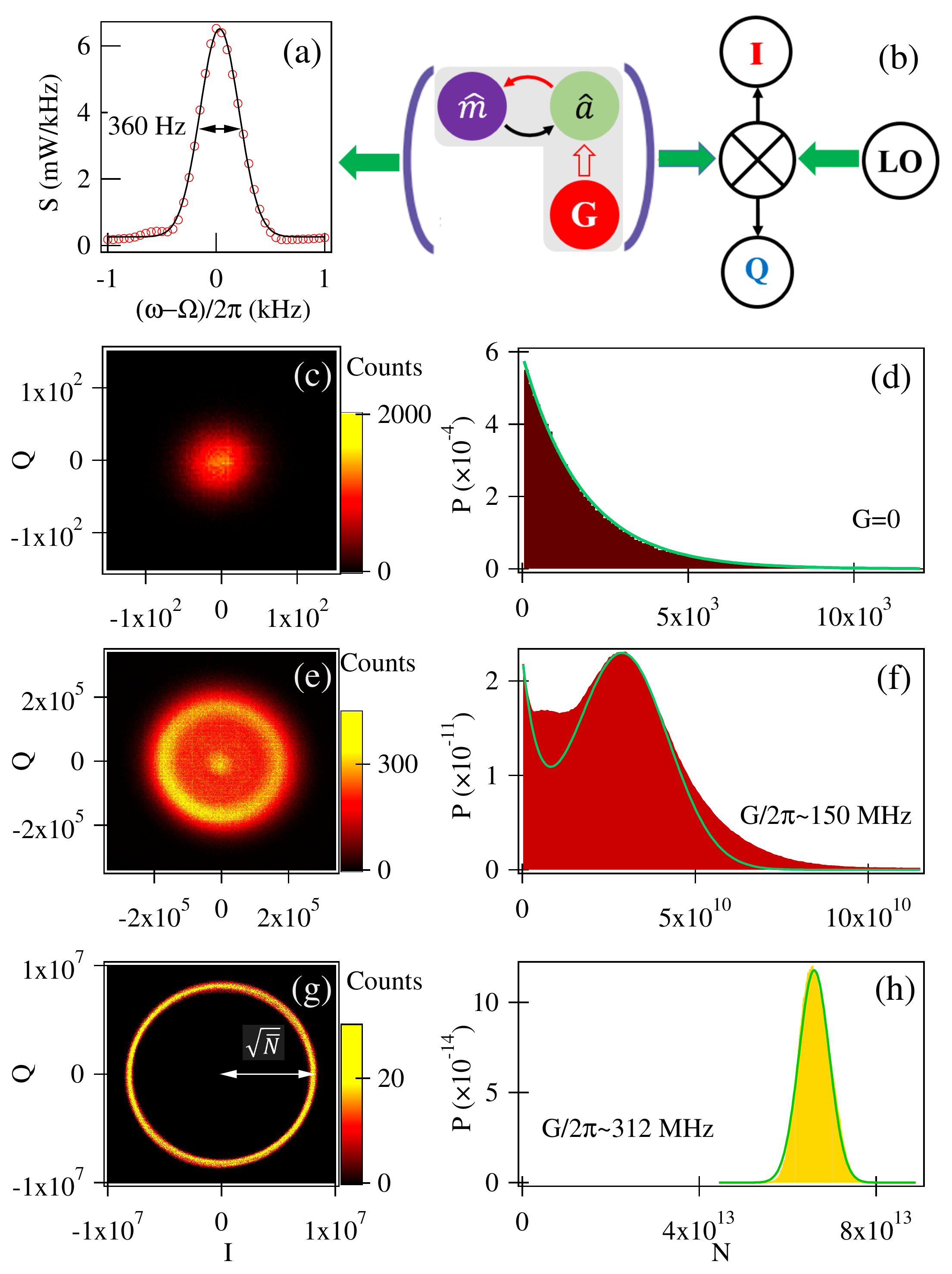}
\caption{\textbf{Coherent emission.}  (a) The power spectral density of the microwave emission measured at $\Omega/2\pi$ = 3.600 GHz. (b) The heterodyne demodulation circuit to perform real-time in-phase (I) and quadrature (Q) measurements. LO represents a local oscillator. The IQ histograms measured by setting the cavity amplifier at (c) 0~V , (e) 0.77~V and (g) 7~V, showing the enhanced amplitude coherence by increasing $G$. The output power ($I^2+Q^2$) is calibrated and converted to the photon number $N$, and the obtained photon number statistics $P(N)$ is fitted to exponential and/or Gaussian distributions in (d),(f), and (h).}
\label{Fig3}
\end{center}
\end{figure}

Emission spectra are measured in the \textquotedblleft memory" setting at $\Delta$ = 0 when up-sweeping. Without any probe signal, we use a signal analyzer to measure the power spectral density $S(f)$ of the emitted photons. As shown in Fig.~\ref{Fig3}(a), an emission peak of about 4.3 mW is detected with $S(f)$ =  6.5 mW/kHz at $\Omega/2\pi$ = 3.600 GHz.  The measured linewidth (full width at half maximum, or FWHM) is 360 Hz, which is remarkable for magnetic systems: via the light-matter interaction, the FWHM is three orders of magnitude smaller than the magnon damping ($\sim$ 900~kHz) that limits the quality factor of spin-toque oscillators \cite{Klselev2003,Kaka2005,Deac2008}. Furthermore, the measured FWHM exceeds the performance of the proposed spin-torque-oscillator maser, where simulations indicate an upper bound emission linewidth of about 4 kHz \cite{Hou2021}.

To quantitatively analyze the photon coherence, we investigate the statistics of the emitted photons. We first mix the emission signal through a modulator board with a local oscillator, and then record the down-converted in-phase (I) and quadrature (Q) (90$^\circ$ out-of-phase) signals with an oscilloscope. The gain-dependence of the amplitude coherence is systematically shown in Fig. \ref{Fig3}. At $V=0$ with $G = 0$,  the IQ histogram in Fig.~\ref{Fig3}(c) shows all outputs around zero. The output power ($I^2+Q^2$) is calibrated \cite{supp} and converted to the photon number $N$, and the obtained photon number distribution $P(N)$ is described as an exponential distribution in Fig. \ref{Fig3}(d), indicating incoherent emission subjective to thermal and charge noises \cite{Liu2015}. Tuning the voltage to 0.77~V ($G \approx \kappa_c$), outside of the center spot of the noise signal, the histogram in Fig. \ref{Fig3}(e) develops non-zero outputs with a donut shape that agrees with the distribution of a coherent source \cite{Liu2015}. Here, $P(N)$ can be approximately fitted by a combined exponential and Gaussian distribution in Fig. \ref{Fig3}(f), where the Gaussian distribution indicates the onset of coherent emission. Setting $V=7$ V ($G \gg \kappa_c$), the IQ histogram vanishes at the center and uniformly distributes in a ring in Fig. \ref{Fig3}(g). The histogram in Fig. \ref{Fig3}(h) can be well fit by the Gaussian lineshape $P(N) \sim e^{(N-\bar{N})^2/2\sigma^2}$ with $\bar{N} = (6.600\pm0.001)\times10^{13}$ and $\sigma = (3.40\pm0.01)\times10^{12}$, demonstrating amplitude coherence of the emitted photons \cite{Cassidy2017}. Furthermore, we have studied the phase coherence of the emission by analyzing the first-order coherence function $g^{(1)} (\tau)$, which constitutes as the convolution of the IQ signal and its complex conjugate with time delay $\tau$. We deduce that at $V=7$ V, the phase coherence time reaches 0.9 ms, confirming the gain-sustained photon coherence of the GDP \cite{supp}.

\begin{figure} [t!]
\begin{center}
\includegraphics[width=\columnwidth]{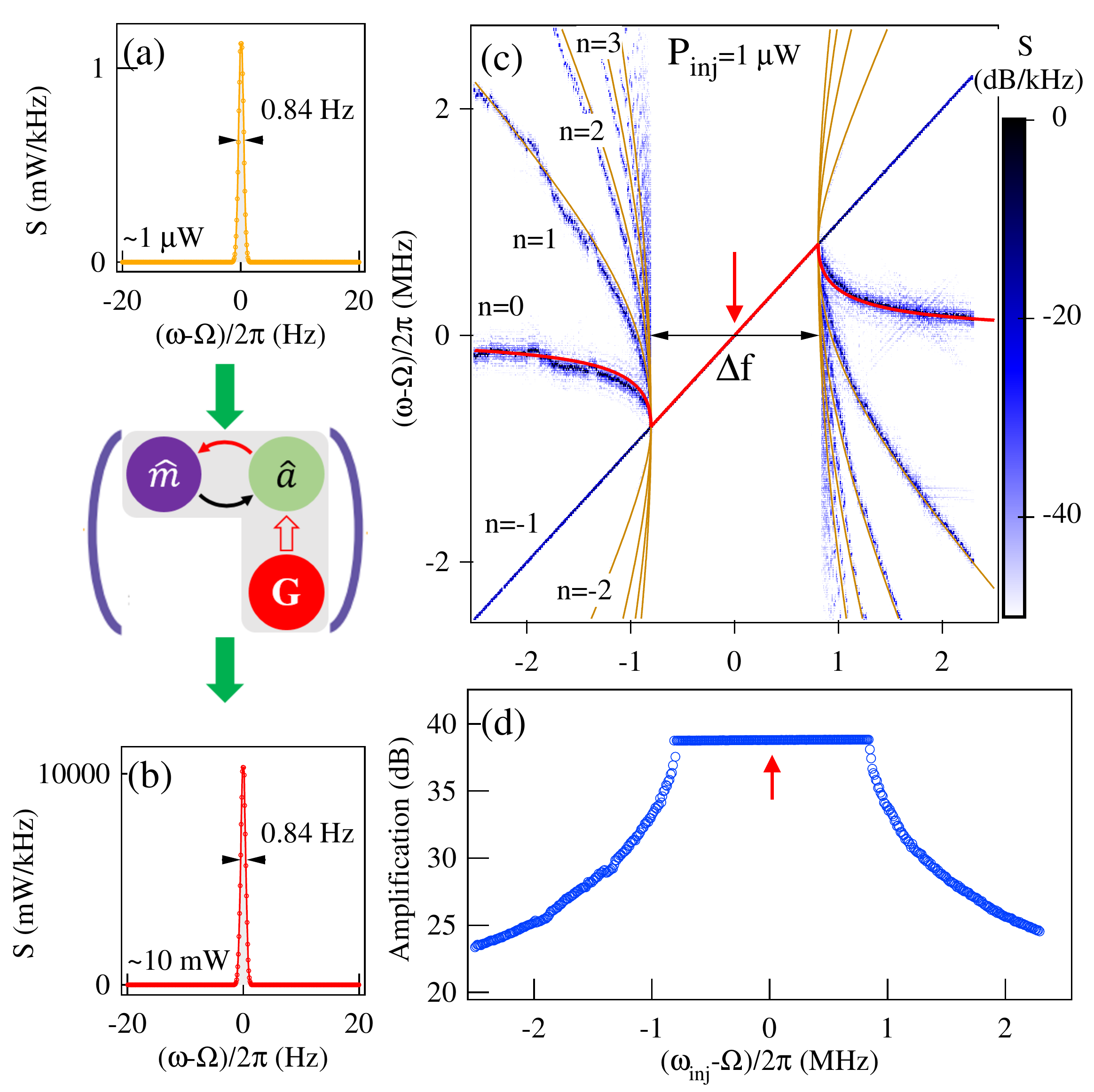}
\caption{\textbf{Coherent amplification.} (a) A ultra-high quality weak microwave signal is injected at $\omega_{\rm inj}/2\pi = \Omega/2\pi$ = 3.594 GHz from a signal generator into the GDP. (b) The output power spectral density $S(f)$ of the photons emitted by the GDP. (c) The $S(f)$ measured by changing the frequency of the injection tone $\omega_{\rm inj}$. The number $n$ indicates the index of the sideband harmonics. The red and orange solid curves are calculated from the theory developed by Adler. (d) Within the injection locking bandwidth of $\Delta f = 1.6$ MHz, the GDP enables about 40 dB coherent amplification [deduced along the diagonal line in (c)].}\label{Fig4}
\end{center}
\end{figure}

Finally, we demonstrate the capability of the GDP for coherently amplifying weak microwave signals. Using a signal generator, we inject the weak signal (about 1 $\mu$W with $1.5\times10^{10}$ photons) into the GDP at the frequency $\omega_{\rm inj}/2\pi = \Omega/2\pi$ = 3.594 GHz. As shown in Fig.~\ref{Fig4}(a), the injected signal has an ultra high-quality (FWHM of 0.84~Hz). With the injected weak signal, we measure the output power spectral density $S(f)$ of the photons emitted by the GDP. As shown in Fig.~\ref{Fig4}(b), the GDP amplifies the input power by 4 orders of magnitude without undermining the ultra-high quality of the input signal.

Why could the GDP, with an intrinsic emission FWHM of about 360 Hz, coherently amplify the signal of the FWHM of 0.84~Hz so well? To clarify the mechanism, we study $S(f)$ by systematically changing $\omega_{\rm inj}$. The measured $S(f)$ mapping is plotted in Fig.~\ref{Fig4}(c), where the solid curves are calculated using Adler's classical theory describing the locking phenomena of oscillators \cite{Adler1946, supp}. It reveals that as an auto-oscillator, the GDP can be phase locked by the weak signal via the injection locking mechanism \cite{Adler1946}. At $|\omega_{\rm inj}-\Omega|/2\pi > 0.8$ MHz, the wave-mixing between the injected signal and the GDP emission produces a frequency comb with sideband oscillations. When $\omega_{\rm inj}$ approaches $\Omega$, the injected signal is amplified, so that all sideband oscillations are suppressed and the frequency of the GDP emission is pulled towards the injection tone. Within the injection locking bandwidth of $\Delta f = 1.6$ MHz, the GDP enables about 40 dB coherent amplification as shown in Fig.~\ref{Fig4}(d). Such a mode-selective amplification boosts GDP's performance as a gain-driven coherent photon source, enhancing its emission quality factor to exceed 10$^9$. The superb emission and amplification performance can be further improved by utilizing the flexible controllability of the cavity magnonics platform. For example, the bandwidth can be easily improved by using frequency-tunable planar cavities.

In summary, we have explored the physics of gain-driven polariton. Contrary to conventional polaritons formed in dissipative systems, here, the light-matter interaction is activated by an amplified electromagnetic field and stabilized by nonlinear damping. A GDP at microwave frequencies is created by incorporating the gain mechanism into the cavity magnonics platform. First, the transmission spectra of GDP are measured, revealing intriguing effects such as the initial condition dependent phase singularity, and spontaneous mode selection that governs magnon-photon synchronization. Then, the photon emission of GDP is studied by analyzing the photon statistics, demonstrating gain-sustained phase and amplitude coherence. By using the injection locking technique, the emission linewidth of the GDP is optimized to below 1~Hz, exceeding many state-of-the-art masers. Our work opens new horizons for exploring gain-driven light-matter interaction. The polariton auto-oscillation revealed in our work may broadly impact the study of synchronization, frequency combs, on-chip masers, and chaotic behaviours in light-matter interactions.

\begin{acknowledgments}

This work has been funded by NSERC Discovery Grants and NSERC Discovery Accelerator Supplements (C.-M. H.).

\end{acknowledgments}


\begin{thebibliography}{42}%
\makeatletter
\providecommand \@ifxundefined [1]{%
 \@ifx{#1\undefined}
}%
\providecommand \@ifnum [1]{%
 \ifnum #1\expandafter \@firstoftwo
 \else \expandafter \@secondoftwo
 \fi
}%
\providecommand \@ifx [1]{%
 \ifx #1\expandafter \@firstoftwo
 \else \expandafter \@secondoftwo
 \fi
}%
\providecommand \natexlab [1]{#1}%
\providecommand \enquote  [1]{``#1''}%
\providecommand \bibnamefont  [1]{#1}%
\providecommand \bibfnamefont [1]{#1}%
\providecommand \citenamefont [1]{#1}%
\providecommand \href@noop [0]{\@secondoftwo}%
\providecommand \href [0]{\begingroup \@sanitize@url \@href}%
\providecommand \@href[1]{\@@startlink{#1}\@@href}%
\providecommand \@@href[1]{\endgroup#1\@@endlink}%
\providecommand \@sanitize@url [0]{\catcode `\\12\catcode `\$12\catcode
  `\&12\catcode `\#12\catcode `\^12\catcode `\_12\catcode `\%12\relax}%
\providecommand \@@startlink[1]{}%
\providecommand \@@endlink[0]{}%
\providecommand \url  [0]{\begingroup\@sanitize@url \@url }%
\providecommand \@url [1]{\endgroup\@href {#1}{\urlprefix }}%
\providecommand \urlprefix  [0]{URL }%
\providecommand \Eprint [0]{\href }%
\providecommand \doibase [0]{http://dx.doi.org/}%
\providecommand \selectlanguage [0]{\@gobble}%
\providecommand \bibinfo  [0]{\@secondoftwo}%
\providecommand \bibfield  [0]{\@secondoftwo}%
\providecommand \translation [1]{[#1]}%
\providecommand \BibitemOpen [0]{}%
\providecommand \bibitemStop [0]{}%
\providecommand \bibitemNoStop [0]{.\EOS\space}%
\providecommand \EOS [0]{\spacefactor3000\relax}%
\providecommand \BibitemShut  [1]{\csname bibitem#1\endcsname}%
\let\auto@bib@innerbib\@empty
\bibitem [{\citenamefont {Artman}\ and\ \citenamefont
  {Tannenwald}(1953)}]{Artman1953}%
  \BibitemOpen
  \bibfield  {author} {\bibinfo {author} {\bibfnamefont {J.~O.}\ \bibnamefont
  {Artman}}\ and\ \bibinfo {author} {\bibfnamefont {P.~E.}\ \bibnamefont
  {Tannenwald}},\ }\href {\doibase 10.1103/PhysRev.91.1014.3} {\bibfield
  {journal} {\bibinfo  {journal} {Physical Review}\ }\textbf {\bibinfo {volume}
  {91}},\ \bibinfo {pages} {1014} (\bibinfo {year} {1953})}\BibitemShut
  {NoStop}%
\bibitem [{\citenamefont {Soykal}\ and\ \citenamefont
  {Flatt{\'{e}}}(2010)}]{Soykal2010}%
  \BibitemOpen
  \bibfield  {author} {\bibinfo {author} {\bibfnamefont {{\"{O}}.~O.}\
  \bibnamefont {Soykal}}\ and\ \bibinfo {author} {\bibfnamefont {M.~E.}\
  \bibnamefont {Flatt{\'{e}}}},\ }\href {\doibase
  10.1103/PhysRevLett.104.077202} {\bibfield  {journal} {\bibinfo  {journal}
  {Physical Review Letters}\ }\textbf {\bibinfo {volume} {104}},\ \bibinfo
  {pages} {077202} (\bibinfo {year} {2010})}\BibitemShut {NoStop}%
\bibitem [{\citenamefont {Huebl}\ \emph {et~al.}(2013)\citenamefont {Huebl},
  \citenamefont {Zollitsch}, \citenamefont {Lotze}, \citenamefont {Hocke},
  \citenamefont {Greifenstein}, \citenamefont {Marx}, \citenamefont {Gross},\
  and\ \citenamefont {Goennenwein}}]{Huebl2013}%
  \BibitemOpen
  \bibfield  {author} {\bibinfo {author} {\bibfnamefont {H.}~\bibnamefont
  {Huebl}}, \bibinfo {author} {\bibfnamefont {C.~W.}\ \bibnamefont
  {Zollitsch}}, \bibinfo {author} {\bibfnamefont {J.}~\bibnamefont {Lotze}},
  \bibinfo {author} {\bibfnamefont {F.}~\bibnamefont {Hocke}}, \bibinfo
  {author} {\bibfnamefont {M.}~\bibnamefont {Greifenstein}}, \bibinfo {author}
  {\bibfnamefont {A.}~\bibnamefont {Marx}}, \bibinfo {author} {\bibfnamefont
  {R.}~\bibnamefont {Gross}}, \ and\ \bibinfo {author} {\bibfnamefont
  {S.~T.~B.}\ \bibnamefont {Goennenwein}},\ }\href {\doibase
  10.1103/PhysRevLett.111.127003} {\bibfield  {journal} {\bibinfo  {journal}
  {Physical Review Letters}\ }\textbf {\bibinfo {volume} {111}},\ \bibinfo
  {pages} {127003} (\bibinfo {year} {2013})}\BibitemShut {NoStop}%
\bibitem [{\citenamefont {Tabuchi}\ \emph {et~al.}(2014)\citenamefont
  {Tabuchi}, \citenamefont {Ishino}, \citenamefont {Ishikawa}, \citenamefont
  {Yamazaki}, \citenamefont {Usami},\ and\ \citenamefont
  {Nakamura}}]{Tabuchi2014}%
  \BibitemOpen
  \bibfield  {author} {\bibinfo {author} {\bibfnamefont {Y.}~\bibnamefont
  {Tabuchi}}, \bibinfo {author} {\bibfnamefont {S.}~\bibnamefont {Ishino}},
  \bibinfo {author} {\bibfnamefont {T.}~\bibnamefont {Ishikawa}}, \bibinfo
  {author} {\bibfnamefont {R.}~\bibnamefont {Yamazaki}}, \bibinfo {author}
  {\bibfnamefont {K.}~\bibnamefont {Usami}}, \ and\ \bibinfo {author}
  {\bibfnamefont {Y.}~\bibnamefont {Nakamura}},\ }\href {\doibase
  10.1103/PhysRevLett.113.083603} {\bibfield  {journal} {\bibinfo  {journal}
  {Physical Review Letters}\ }\textbf {\bibinfo {volume} {113}},\ \bibinfo
  {pages} {083603} (\bibinfo {year} {2014})}\BibitemShut {NoStop}%
\bibitem [{\citenamefont {Zhang}\ \emph {et~al.}(2014)\citenamefont {Zhang},
  \citenamefont {Zou}, \citenamefont {Jiang},\ and\ \citenamefont
  {Tang}}]{Zhang2014}%
  \BibitemOpen
  \bibfield  {author} {\bibinfo {author} {\bibfnamefont {X.}~\bibnamefont
  {Zhang}}, \bibinfo {author} {\bibfnamefont {C.-L.}\ \bibnamefont {Zou}},
  \bibinfo {author} {\bibfnamefont {L.}~\bibnamefont {Jiang}}, \ and\ \bibinfo
  {author} {\bibfnamefont {H.~X.}\ \bibnamefont {Tang}},\ }\href {\doibase
  10.1103/PhysRevLett.113.156401} {\bibfield  {journal} {\bibinfo  {journal}
  {Physical Review Letters}\ }\textbf {\bibinfo {volume} {113}},\ \bibinfo
  {pages} {156401} (\bibinfo {year} {2014})}\BibitemShut {NoStop}%
\bibitem [{\citenamefont {Bai}\ \emph {et~al.}(2015)\citenamefont {Bai},
		\citenamefont {Harder}, \citenamefont {Chen}, \citenamefont {Fan},
		\citenamefont {Xiao},\ and\ \citenamefont {Hu}}]{Bai2015}%
	\BibitemOpen
	\bibfield  {author} {\bibinfo {author} {\bibfnamefont {L.}~\bibnamefont
			{Bai}}, \bibinfo {author} {\bibfnamefont {M.}~\bibnamefont {Harder}},
		\bibinfo {author} {\bibfnamefont {Y.~P.}\ \bibnamefont {Chen}}, \bibinfo
		{author} {\bibfnamefont {X.}~\bibnamefont {Fan}}, \bibinfo {author}
		{\bibfnamefont {J.~Q.}\ \bibnamefont {Xiao}}, \ and\ \bibinfo {author}
		{\bibfnamefont {C.-M.}\ \bibnamefont {Hu}},\ }\href {\doibase
  10.1103/PhysRevLett.114.227201} {\bibfield  {journal} {\bibinfo  {journal}
  {Physical Review Letters}\ }\textbf {\bibinfo {volume} {114}},\ \bibinfo
  {pages} {227201} (\bibinfo {year} {2015})}\BibitemShut {NoStop}%
\bibitem [{\citenamefont {{Zare Rameshti}}\ \emph {et~al.}(2022)\citenamefont
  {{Zare Rameshti}}, \citenamefont {{Viola Kusminskiy}}, \citenamefont {Haigh},
  \citenamefont {Usami}, \citenamefont {Lachance-Quirion}, \citenamefont
  {Nakamura}, \citenamefont {Hu}, \citenamefont {Tang}, \citenamefont {Bauer},\
  and\ \citenamefont {Blanter}}]{Rameshti2021}%
  \BibitemOpen
  \bibfield  {author} {\bibinfo {author} {\bibfnamefont {B.}~\bibnamefont
  {{Zare Rameshti}}}, \bibinfo {author} {\bibfnamefont {S.}~\bibnamefont
  {{Viola Kusminskiy}}}, \bibinfo {author} {\bibfnamefont {J.~A.}\ \bibnamefont
  {Haigh}}, \bibinfo {author} {\bibfnamefont {K.}~\bibnamefont {Usami}},
  \bibinfo {author} {\bibfnamefont {D.}~\bibnamefont {Lachance-Quirion}},
  \bibinfo {author} {\bibfnamefont {Y.}~\bibnamefont {Nakamura}}, \bibinfo
  {author} {\bibfnamefont {C.-M.}\ \bibnamefont {Hu}}, \bibinfo {author}
  {\bibfnamefont {H.~X.}\ \bibnamefont {Tang}}, \bibinfo {author}
  {\bibfnamefont {G.~E.}\ \bibnamefont {Bauer}}, \ and\ \bibinfo {author}
  {\bibfnamefont {Y.~M.}\ \bibnamefont {Blanter}},\ }\href {\doibase
  10.1016/j.physrep.2022.06.001} {\bibfield  {journal} {\bibinfo  {journal}
  {Physics Reports}\ }\textbf {\bibinfo {volume} {979}},\ \bibinfo {pages} {1}
  (\bibinfo {year} {2022})}\BibitemShut {NoStop}%
\bibitem [{\citenamefont {Lachance-Quirion}\ \emph {et~al.}(2019)\citenamefont
  {Lachance-Quirion}, \citenamefont {Tabuchi}, \citenamefont {Gloppe},
  \citenamefont {Usami},\ and\ \citenamefont
  {Nakamura}}]{Lachance-Quirion2019}%
  \BibitemOpen
  \bibfield  {author} {\bibinfo {author} {\bibfnamefont {D.}~\bibnamefont
  {Lachance-Quirion}}, \bibinfo {author} {\bibfnamefont {Y.}~\bibnamefont
  {Tabuchi}}, \bibinfo {author} {\bibfnamefont {A.}~\bibnamefont {Gloppe}},
  \bibinfo {author} {\bibfnamefont {K.}~\bibnamefont {Usami}}, \ and\ \bibinfo
  {author} {\bibfnamefont {Y.}~\bibnamefont {Nakamura}},\ }\href {\doibase
  10.7567/1882-0786/ab248d} {\bibfield  {journal} {\bibinfo  {journal} {Applied
  Physics Express}\ }\textbf {\bibinfo {volume} {12}},\ \bibinfo {pages}
  {070101} (\bibinfo {year} {2019})}\BibitemShut {NoStop}%
\bibitem [{\citenamefont {Zhang}\ \emph {et~al.}(2019)\citenamefont {Zhang},
  \citenamefont {Scully},\ and\ \citenamefont {Agarwal}}]{Zhang2019}%
  \BibitemOpen
  \bibfield  {author} {\bibinfo {author} {\bibfnamefont {Z.}~\bibnamefont
  {Zhang}}, \bibinfo {author} {\bibfnamefont {M.~O.}\ \bibnamefont {Scully}}, \
  and\ \bibinfo {author} {\bibfnamefont {G.~S.}\ \bibnamefont {Agarwal}},\
  }\href {\doibase 10.1103/PhysRevResearch.1.023021} {\bibfield  {journal}
  {\bibinfo  {journal} {Physical Review Research}\ }\textbf {\bibinfo {volume}
  {1}},\ \bibinfo {pages} {023021} (\bibinfo {year} {2019})}\BibitemShut
  {NoStop}%
\bibitem [{\citenamefont {Clerk}\ \emph {et~al.}(2020)\citenamefont {Clerk},
  \citenamefont {Lehnert}, \citenamefont {Bertet}, \citenamefont {Petta},\ and\
  \citenamefont {Nakamura}}]{Clerk2020}%
  \BibitemOpen
  \bibfield  {author} {\bibinfo {author} {\bibfnamefont {A.~A.}\ \bibnamefont
  {Clerk}}, \bibinfo {author} {\bibfnamefont {K.~W.}\ \bibnamefont {Lehnert}},
  \bibinfo {author} {\bibfnamefont {P.}~\bibnamefont {Bertet}}, \bibinfo
  {author} {\bibfnamefont {J.~R.}\ \bibnamefont {Petta}}, \ and\ \bibinfo
  {author} {\bibfnamefont {Y.}~\bibnamefont {Nakamura}},\ }\href {\doibase
  10.1038/s41567-020-0797-9} {\bibfield  {journal} {\bibinfo  {journal} {Nature
  Physics}\ }\textbf {\bibinfo {volume} {16}},\ \bibinfo {pages} {257}
  (\bibinfo {year} {2020})}\BibitemShut {NoStop}%
\bibitem [{\citenamefont {Zhang}\ \emph {et~al.}(2016)\citenamefont {Zhang},
  \citenamefont {Zou}, \citenamefont {Jiang},\ and\ \citenamefont
  {Tang}}]{Zhang2016}%
  \BibitemOpen
  \bibfield  {author} {\bibinfo {author} {\bibfnamefont {X.}~\bibnamefont
  {Zhang}}, \bibinfo {author} {\bibfnamefont {C.~L.}\ \bibnamefont {Zou}},
  \bibinfo {author} {\bibfnamefont {L.}~\bibnamefont {Jiang}}, \ and\ \bibinfo
  {author} {\bibfnamefont {H.~X.}\ \bibnamefont {Tang}},\ }\href {\doibase
  10.1126/sciadv.1501286} {\bibfield  {journal} {\bibinfo  {journal} {Science
  Advances}\ }\textbf {\bibinfo {volume} {2}},\ \bibinfo {pages} {e1501286}
  (\bibinfo {year} {2016})}\BibitemShut {NoStop}%
\bibitem [{\citenamefont {Osada}\ \emph {et~al.}(2016)\citenamefont {Osada},
  \citenamefont {Hisatomi}, \citenamefont {Noguchi}, \citenamefont {Tabuchi},
  \citenamefont {Yamazaki}, \citenamefont {Usami}, \citenamefont {Sadgrove},
  \citenamefont {Yalla}, \citenamefont {Nomura},\ and\ \citenamefont
  {Nakamura}}]{Osada2016}%
  \BibitemOpen
  \bibfield  {author} {\bibinfo {author} {\bibfnamefont {A.}~\bibnamefont
  {Osada}}, \bibinfo {author} {\bibfnamefont {R.}~\bibnamefont {Hisatomi}},
  \bibinfo {author} {\bibfnamefont {A.}~\bibnamefont {Noguchi}}, \bibinfo
  {author} {\bibfnamefont {Y.}~\bibnamefont {Tabuchi}}, \bibinfo {author}
  {\bibfnamefont {R.}~\bibnamefont {Yamazaki}}, \bibinfo {author}
  {\bibfnamefont {K.}~\bibnamefont {Usami}}, \bibinfo {author} {\bibfnamefont
  {M.}~\bibnamefont {Sadgrove}}, \bibinfo {author} {\bibfnamefont
  {R.}~\bibnamefont {Yalla}}, \bibinfo {author} {\bibfnamefont
  {M.}~\bibnamefont {Nomura}}, \ and\ \bibinfo {author} {\bibfnamefont
  {Y.}~\bibnamefont {Nakamura}},\ }\href {\doibase
  10.1103/PhysRevLett.116.223601} {\bibfield  {journal} {\bibinfo  {journal}
  {Physical Review Letters}\ }\textbf {\bibinfo {volume} {116}},\ \bibinfo
  {pages} {223601} (\bibinfo {year} {2016})}\BibitemShut {NoStop}%
\bibitem [{\citenamefont {Potts}\ \emph {et~al.}(2021)\citenamefont {Potts},
  \citenamefont {Varga}, \citenamefont {Bittencourt}, \citenamefont
  {Kusminskiy},\ and\ \citenamefont {Davis}}]{Potts2021}%
  \BibitemOpen
  \bibfield  {author} {\bibinfo {author} {\bibfnamefont {C.~A.}\ \bibnamefont
  {Potts}}, \bibinfo {author} {\bibfnamefont {E.}~\bibnamefont {Varga}},
  \bibinfo {author} {\bibfnamefont {V.~A. S.~V.}\ \bibnamefont {Bittencourt}},
  \bibinfo {author} {\bibfnamefont {S.~V.}\ \bibnamefont {Kusminskiy}}, \ and\
  \bibinfo {author} {\bibfnamefont {J.~P.}\ \bibnamefont {Davis}},\ }\href
  {\doibase 10.1103/PhysRevX.11.031053} {\bibfield  {journal} {\bibinfo
  {journal} {Physical Review X}\ }\textbf {\bibinfo {volume} {11}},\ \bibinfo
  {pages} {031053} (\bibinfo {year} {2021})}\BibitemShut {NoStop}%
\bibitem [{\citenamefont {Zhang}\ \emph {et~al.}(2015)\citenamefont {Zhang},
  \citenamefont {Zou}, \citenamefont {Zhu}, \citenamefont {Marquardt},
  \citenamefont {Jiang},\ and\ \citenamefont {Tang}}]{Zhang2015a}%
  \BibitemOpen
  \bibfield  {author} {\bibinfo {author} {\bibfnamefont {X.}~\bibnamefont
  {Zhang}}, \bibinfo {author} {\bibfnamefont {C.-l.}\ \bibnamefont {Zou}},
  \bibinfo {author} {\bibfnamefont {N.}~\bibnamefont {Zhu}}, \bibinfo {author}
  {\bibfnamefont {F.}~\bibnamefont {Marquardt}}, \bibinfo {author}
  {\bibfnamefont {L.}~\bibnamefont {Jiang}}, \ and\ \bibinfo {author}
  {\bibfnamefont {H.~X.}\ \bibnamefont {Tang}},\ }\href {\doibase
  10.1038/ncomms9914} {\bibfield  {journal} {\bibinfo  {journal} {Nature
  Communications}\ }\textbf {\bibinfo {volume} {6}},\ \bibinfo {pages} {8914}
  (\bibinfo {year} {2015})}\BibitemShut {NoStop}%
\bibitem [{\citenamefont {Shen}\ \emph {et~al.}(2021)\citenamefont {Shen},
  \citenamefont {Wang}, \citenamefont {Li}, \citenamefont {Zhu}, \citenamefont
  {Agarwal},\ and\ \citenamefont {You}}]{Shen2021}%
  \BibitemOpen
  \bibfield  {author} {\bibinfo {author} {\bibfnamefont {R.-C.}\ \bibnamefont
  {Shen}}, \bibinfo {author} {\bibfnamefont {Y.-P.}\ \bibnamefont {Wang}},
  \bibinfo {author} {\bibfnamefont {J.}~\bibnamefont {Li}}, \bibinfo {author}
  {\bibfnamefont {S.-Y.}\ \bibnamefont {Zhu}}, \bibinfo {author} {\bibfnamefont
  {G.~S.}\ \bibnamefont {Agarwal}}, \ and\ \bibinfo {author} {\bibfnamefont
  {J.~Q.}\ \bibnamefont {You}},\ }\href {\doibase
  10.1103/PhysRevLett.127.183202} {\bibfield  {journal} {\bibinfo  {journal}
  {Physical Review Letters}\ }\textbf {\bibinfo {volume} {127}},\ \bibinfo
  {pages} {183202} (\bibinfo {year} {2021})}\BibitemShut {NoStop}%
\bibitem [{\citenamefont {Bai}\ \emph {et~al.}(2017)\citenamefont {Bai},
  \citenamefont {Harder}, \citenamefont {Hyde}, \citenamefont {Zhang},
  \citenamefont {Hu}, \citenamefont {Chen},\ and\ \citenamefont
  {Xiao}}]{Bai2017}%
  \BibitemOpen
  \bibfield  {author} {\bibinfo {author} {\bibfnamefont {L.}~\bibnamefont
  {Bai}}, \bibinfo {author} {\bibfnamefont {M.}~\bibnamefont {Harder}},
  \bibinfo {author} {\bibfnamefont {P.}~\bibnamefont {Hyde}}, \bibinfo {author}
  {\bibfnamefont {Z.}~\bibnamefont {Zhang}}, \bibinfo {author} {\bibfnamefont
  {C.-M.}\ \bibnamefont {Hu}}, \bibinfo {author} {\bibfnamefont
  {Y.}~\bibnamefont {Chen}}, \ and\ \bibinfo {author} {\bibfnamefont {J.~Q.}\
  \bibnamefont {Xiao}},\ }\href {\doibase 10.1103/PhysRevLett.118.217201}
  {\bibfield  {journal} {\bibinfo  {journal} {Physical Review Letters}\
  }\textbf {\bibinfo {volume} {118}},\ \bibinfo {pages} {217201} (\bibinfo
  {year} {2017})}\BibitemShut {NoStop}%
\bibitem [{\citenamefont {Lachance-Quirion}\ \emph {et~al.}(2020)\citenamefont
  {Lachance-Quirion}, \citenamefont {Wolski}, \citenamefont {Tabuchi},
  \citenamefont {Kono}, \citenamefont {Usami},\ and\ \citenamefont
  {Nakamura}}]{Lachance-Quirion2020}%
  \BibitemOpen
  \bibfield  {author} {\bibinfo {author} {\bibfnamefont {D.}~\bibnamefont
  {Lachance-Quirion}}, \bibinfo {author} {\bibfnamefont {S.~P.}\ \bibnamefont
  {Wolski}}, \bibinfo {author} {\bibfnamefont {Y.}~\bibnamefont {Tabuchi}},
  \bibinfo {author} {\bibfnamefont {S.}~\bibnamefont {Kono}}, \bibinfo {author}
  {\bibfnamefont {K.}~\bibnamefont {Usami}}, \ and\ \bibinfo {author}
  {\bibfnamefont {Y.}~\bibnamefont {Nakamura}},\ }\href {\doibase
  10.1126/science.aaz9236} {\bibfield  {journal} {\bibinfo  {journal}
  {Science}\ }\textbf {\bibinfo {volume} {367}},\ \bibinfo {pages} {425}
  (\bibinfo {year} {2020})}\BibitemShut {NoStop}%
\bibitem [{\citenamefont {Haigh}\ \emph {et~al.}(2016)\citenamefont {Haigh},
  \citenamefont {Nunnenkamp}, \citenamefont {Ramsay},\ and\ \citenamefont
  {Ferguson}}]{Haigh2016}%
  \BibitemOpen
  \bibfield  {author} {\bibinfo {author} {\bibfnamefont {J.~A.}\ \bibnamefont
  {Haigh}}, \bibinfo {author} {\bibfnamefont {A.}~\bibnamefont {Nunnenkamp}},
  \bibinfo {author} {\bibfnamefont {A.~J.}\ \bibnamefont {Ramsay}}, \ and\
  \bibinfo {author} {\bibfnamefont {A.~J.}\ \bibnamefont {Ferguson}},\ }\href
  {\doibase 10.1103/PhysRevLett.117.133602} {\bibfield  {journal} {\bibinfo
  {journal} {Physical Review Letters}\ }\textbf {\bibinfo {volume} {117}},\
  \bibinfo {pages} {133602} (\bibinfo {year} {2016})}\BibitemShut {NoStop}%
\bibitem [{\citenamefont {Wang}\ \emph {et~al.}(2018)\citenamefont {Wang},
  \citenamefont {Zhang}, \citenamefont {Zhang}, \citenamefont {Li},
  \citenamefont {Hu},\ and\ \citenamefont {You}}]{Wang2018}%
  \BibitemOpen
  \bibfield  {author} {\bibinfo {author} {\bibfnamefont {Y.-P.}\ \bibnamefont
  {Wang}}, \bibinfo {author} {\bibfnamefont {G.-Q.}\ \bibnamefont {Zhang}},
  \bibinfo {author} {\bibfnamefont {D.}~\bibnamefont {Zhang}}, \bibinfo
  {author} {\bibfnamefont {T.-F.}\ \bibnamefont {Li}}, \bibinfo {author}
  {\bibfnamefont {C.-M.}\ \bibnamefont {Hu}}, \ and\ \bibinfo {author}
  {\bibfnamefont {J.~Q.}\ \bibnamefont {You}},\ }\href {\doibase
  10.1103/PhysRevLett.120.057202} {\bibfield  {journal} {\bibinfo  {journal}
  {Physical Review Letters}\ }\textbf {\bibinfo {volume} {120}},\ \bibinfo
  {pages} {057202} (\bibinfo {year} {2018})}\BibitemShut {NoStop}%
\bibitem [{\citenamefont {Harder}\ \emph {et~al.}(2018)\citenamefont {Harder},
  \citenamefont {Yang}, \citenamefont {Yao}, \citenamefont {Yu}, \citenamefont
  {Rao}, \citenamefont {Gui}, \citenamefont {Stamps},\ and\ \citenamefont
  {Hu}}]{Harder2018}%
  \BibitemOpen
  \bibfield  {author} {\bibinfo {author} {\bibfnamefont {M.}~\bibnamefont
  {Harder}}, \bibinfo {author} {\bibfnamefont {Y.}~\bibnamefont {Yang}},
  \bibinfo {author} {\bibfnamefont {B.~M.}\ \bibnamefont {Yao}}, \bibinfo
  {author} {\bibfnamefont {C.~H.}\ \bibnamefont {Yu}}, \bibinfo {author}
  {\bibfnamefont {J.~W.}\ \bibnamefont {Rao}}, \bibinfo {author} {\bibfnamefont
  {Y.~S.}\ \bibnamefont {Gui}}, \bibinfo {author} {\bibfnamefont {R.~L.}\
  \bibnamefont {Stamps}}, \ and\ \bibinfo {author} {\bibfnamefont {C.-M.}\
  \bibnamefont {Hu}},\ }\href {\doibase 10.1103/PhysRevLett.121.137203}
  {\bibfield  {journal} {\bibinfo  {journal} {Physical Review Letters}\
  }\textbf {\bibinfo {volume} {121}},\ \bibinfo {pages} {137203} (\bibinfo
  {year} {2018})}\BibitemShut {NoStop}%
\bibitem [{\citenamefont {Bhoi}\ \emph {et~al.}(2019)\citenamefont {Bhoi},
  \citenamefont {Kim}, \citenamefont {Jang}, \citenamefont {Kim}, \citenamefont
  {Yang}, \citenamefont {Cho},\ and\ \citenamefont {Kim}}]{Bhoi2019}%
  \BibitemOpen
  \bibfield  {author} {\bibinfo {author} {\bibfnamefont {B.}~\bibnamefont
  {Bhoi}}, \bibinfo {author} {\bibfnamefont {B.}~\bibnamefont {Kim}}, \bibinfo
  {author} {\bibfnamefont {S.-h.}\ \bibnamefont {Jang}}, \bibinfo {author}
  {\bibfnamefont {J.}~\bibnamefont {Kim}}, \bibinfo {author} {\bibfnamefont
  {J.}~\bibnamefont {Yang}}, \bibinfo {author} {\bibfnamefont {Y.-J.}\
  \bibnamefont {Cho}}, \ and\ \bibinfo {author} {\bibfnamefont {S.-K.}\
  \bibnamefont {Kim}},\ }\href {\doibase 10.1103/PhysRevB.99.134426} {\bibfield
   {journal} {\bibinfo  {journal} {Physical Review B}\ }\textbf {\bibinfo
  {volume} {99}},\ \bibinfo {pages} {134426} (\bibinfo {year}
  {2019})}\BibitemShut {NoStop}%
\bibitem [{\citenamefont {Boventer}\ \emph {et~al.}(2019)\citenamefont
  {Boventer}, \citenamefont {Kl{\"{a}}ui}, \citenamefont {Mac{\^{e}}do},\ and\
  \citenamefont {Weides}}]{Boventer2019}%
  \BibitemOpen
  \bibfield  {author} {\bibinfo {author} {\bibfnamefont {I.}~\bibnamefont
  {Boventer}}, \bibinfo {author} {\bibfnamefont {M.}~\bibnamefont
  {Kl{\"{a}}ui}}, \bibinfo {author} {\bibfnamefont {R.}~\bibnamefont
  {Mac{\^{e}}do}}, \ and\ \bibinfo {author} {\bibfnamefont {M.}~\bibnamefont
  {Weides}},\ }\href {\doibase 10.1088/1367-2630/ab5c12} {\bibfield  {journal}
  {\bibinfo  {journal} {New Journal of Physics}\ }\textbf {\bibinfo {volume}
  {21}},\ \bibinfo {pages} {125001} (\bibinfo {year} {2019})}\BibitemShut
  {NoStop}%
\bibitem [{\citenamefont {Zhang}\ \emph {et~al.}(2017)\citenamefont {Zhang},
  \citenamefont {Luo}, \citenamefont {Wang}, \citenamefont {Li},\ and\
  \citenamefont {You}}]{Zhang2017}%
  \BibitemOpen
  \bibfield  {author} {\bibinfo {author} {\bibfnamefont {D.}~\bibnamefont
  {Zhang}}, \bibinfo {author} {\bibfnamefont {X.-Q.}\ \bibnamefont {Luo}},
  \bibinfo {author} {\bibfnamefont {Y.-P.}\ \bibnamefont {Wang}}, \bibinfo
  {author} {\bibfnamefont {T.-F.}\ \bibnamefont {Li}}, \ and\ \bibinfo {author}
  {\bibfnamefont {J.~Q.}\ \bibnamefont {You}},\ }\href {\doibase
  10.1038/s41467-017-01634-w} {\bibfield  {journal} {\bibinfo  {journal}
  {Nature Communications}\ }\textbf {\bibinfo {volume} {8}},\ \bibinfo {pages}
  {1368} (\bibinfo {year} {2017})}\BibitemShut {NoStop}%
\bibitem [{\citenamefont {Yao}\ \emph {et~al.}(2017)\citenamefont {Yao},
  \citenamefont {Gui}, \citenamefont {Rao}, \citenamefont {Kaur}, \citenamefont
  {Chen}, \citenamefont {Lu}, \citenamefont {Xiao}, \citenamefont {Guo},
  \citenamefont {Marzlin},\ and\ \citenamefont {Hu}}]{Yao2017}%
  \BibitemOpen
  \bibfield  {author} {\bibinfo {author} {\bibfnamefont {B.~M.}~\bibnamefont
  {Yao}}, \bibinfo {author} {\bibfnamefont {Y.~S.}\ \bibnamefont {Gui}},
  \bibinfo {author} {\bibfnamefont {J.~W.}\ \bibnamefont {Rao}}, \bibinfo
  {author} {\bibfnamefont {S.}~\bibnamefont {Kaur}}, \bibinfo {author}
  {\bibfnamefont {X.~S.}\ \bibnamefont {Chen}}, \bibinfo {author}
  {\bibfnamefont {W.}~\bibnamefont {Lu}}, \bibinfo {author} {\bibfnamefont
  {Y.}~\bibnamefont {Xiao}}, \bibinfo {author} {\bibfnamefont {H.}~\bibnamefont
  {Guo}}, \bibinfo {author} {\bibfnamefont {K.~P.}\ \bibnamefont {Marzlin}}, \
  and\ \bibinfo {author} {\bibfnamefont {C.-M.}\ \bibnamefont {Hu}},\ }\href
  {\doibase 10.1038/s41467-017-01796-7} {\bibfield  {journal} {\bibinfo
  {journal} {Nature Communications}\ }\textbf {\bibinfo {volume} {8}},\
  \bibinfo {pages} {1} (\bibinfo {year} {2017})}\BibitemShut {NoStop}%
\bibitem [{\citenamefont {Xu}\ \emph {et~al.}(2020)\citenamefont {Xu},
  \citenamefont {Zhong}, \citenamefont {Han}, \citenamefont {Jin},
  \citenamefont {Jiang},\ and\ \citenamefont {Zhang}}]{Xu2020}%
  \BibitemOpen
  \bibfield  {author} {\bibinfo {author} {\bibfnamefont {J.}~\bibnamefont
  {Xu}}, \bibinfo {author} {\bibfnamefont {C.}~\bibnamefont {Zhong}}, \bibinfo
  {author} {\bibfnamefont {X.}~\bibnamefont {Han}}, \bibinfo {author}
  {\bibfnamefont {D.}~\bibnamefont {Jin}}, \bibinfo {author} {\bibfnamefont
  {L.}~\bibnamefont {Jiang}}, \ and\ \bibinfo {author} {\bibfnamefont
  {X.}~\bibnamefont {Zhang}},\ }\href {\doibase 10.1103/PhysRevLett.125.237201}
  {\bibfield  {journal} {\bibinfo  {journal} {Physical Review Letters}\
  }\textbf {\bibinfo {volume} {125}},\ \bibinfo {pages} {237201} (\bibinfo
  {year} {2020})}\BibitemShut {NoStop}%
\bibitem [{\citenamefont {Wang}\ \emph {et~al.}(2019)\citenamefont {Wang},
  \citenamefont {Rao}, \citenamefont {Yang}, \citenamefont {Xu}, \citenamefont
  {Gui}, \citenamefont {Yao}, \citenamefont {You},\ and\ \citenamefont
  {Hu}}]{Wang2019}%
  \BibitemOpen
  \bibfield  {author} {\bibinfo {author} {\bibfnamefont {Y.-P.}\ \bibnamefont
  {Wang}}, \bibinfo {author} {\bibfnamefont {J.~W.}\ \bibnamefont {Rao}},
  \bibinfo {author} {\bibfnamefont {Y.}~\bibnamefont {Yang}}, \bibinfo {author}
  {\bibfnamefont {P.-C.}\ \bibnamefont {Xu}}, \bibinfo {author} {\bibfnamefont
  {Y.~S.}\ \bibnamefont {Gui}}, \bibinfo {author} {\bibfnamefont {B.~M.}\
  \bibnamefont {Yao}}, \bibinfo {author} {\bibfnamefont {J.~Q.}\ \bibnamefont
  {You}}, \ and\ \bibinfo {author} {\bibfnamefont {C.-M.}\ \bibnamefont {Hu}},\
  }\href {\doibase 10.1103/PhysRevLett.123.127202} {\bibfield  {journal}
  {\bibinfo  {journal} {Physical Review Letters}\ }\textbf {\bibinfo {volume}
  {123}},\ \bibinfo {pages} {127202} (\bibinfo {year} {2019})}\BibitemShut
  {NoStop}%
\bibitem [{\citenamefont {Yang}\ \emph {et~al.}(2020)\citenamefont {Yang},
  \citenamefont {Wang}, \citenamefont {Rao}, \citenamefont {Gui}, \citenamefont
  {Yao}, \citenamefont {Lu},\ and\ \citenamefont {Hu}}]{Yang2020}%
  \BibitemOpen
  \bibfield  {author} {\bibinfo {author} {\bibfnamefont {Y.}~\bibnamefont
  {Yang}}, \bibinfo {author} {\bibfnamefont {Y.-P.}\ \bibnamefont {Wang}},
  \bibinfo {author} {\bibfnamefont {J.~W.}\ \bibnamefont {Rao}}, \bibinfo
  {author} {\bibfnamefont {Y.~S.}\ \bibnamefont {Gui}}, \bibinfo {author}
  {\bibfnamefont {B.~M.}\ \bibnamefont {Yao}}, \bibinfo {author} {\bibfnamefont
  {W.}~\bibnamefont {Lu}}, \ and\ \bibinfo {author} {\bibfnamefont {C.-M.}\
  \bibnamefont {Hu}},\ }\href {\doibase 10.1103/PhysRevLett.125.147202}
  {\bibfield  {journal} {\bibinfo  {journal} {Physical Review Letters}\
  }\textbf {\bibinfo {volume} {125}},\ \bibinfo {pages} {147202} (\bibinfo
  {year} {2020})}\BibitemShut {NoStop}%
\bibitem [{\citenamefont {Deng}\ \emph {et~al.}(2002)\citenamefont {Deng},
  \citenamefont {Weihs}, \citenamefont {Santori}, \citenamefont {Bloch},\ and\
  \citenamefont {Yamamoto}}]{Deng2002}%
  \BibitemOpen
  \bibfield  {author} {\bibinfo {author} {\bibfnamefont {H.}~\bibnamefont
  {Deng}}, \bibinfo {author} {\bibfnamefont {G.}~\bibnamefont {Weihs}},
  \bibinfo {author} {\bibfnamefont {C.}~\bibnamefont {Santori}}, \bibinfo
  {author} {\bibfnamefont {J.}~\bibnamefont {Bloch}}, \ and\ \bibinfo {author}
  {\bibfnamefont {Y.}~\bibnamefont {Yamamoto}},\ }\href {\doibase
  10.1126/science.1074464} {\bibfield  {journal} {\bibinfo  {journal}
  {Science}\ }\textbf {\bibinfo {volume} {298}},\ \bibinfo {pages} {199}
  (\bibinfo {year} {2002})}\BibitemShut {NoStop}%
\bibitem [{\citenamefont {Kasprzak}\ \emph {et~al.}(2006)\citenamefont
  {Kasprzak}, \citenamefont {Richard}, \citenamefont {Kundermann},
  \citenamefont {Baas}, \citenamefont {Jeambrun}, \citenamefont {Keeling},
  \citenamefont {Marchetti}, \citenamefont {Szyma{\'{n}}ska}, \citenamefont
  {Andr{\'{e}}}, \citenamefont {Staehli}, \citenamefont {Savona}, \citenamefont
  {Littlewood}, \citenamefont {Deveaud},\ and\ \citenamefont
  {Dang}}]{Kasprzak2006}%
  \BibitemOpen
  \bibfield  {author} {\bibinfo {author} {\bibfnamefont {J.}~\bibnamefont
  {Kasprzak}}, \bibinfo {author} {\bibfnamefont {M.}~\bibnamefont {Richard}},
  \bibinfo {author} {\bibfnamefont {S.}~\bibnamefont {Kundermann}}, \bibinfo
  {author} {\bibfnamefont {A.}~\bibnamefont {Baas}}, \bibinfo {author}
  {\bibfnamefont {P.}~\bibnamefont {Jeambrun}}, \bibinfo {author}
  {\bibfnamefont {J.~M.~J.}\ \bibnamefont {Keeling}}, \bibinfo {author}
  {\bibfnamefont {F.~M.}\ \bibnamefont {Marchetti}}, \bibinfo {author}
  {\bibfnamefont {M.~H.}\ \bibnamefont {Szyma{\'{n}}ska}}, \bibinfo {author}
  {\bibfnamefont {R.}~\bibnamefont {Andr{\'{e}}}}, \bibinfo {author}
  {\bibfnamefont {J.~L.}\ \bibnamefont {Staehli}}, \bibinfo {author}
  {\bibfnamefont {V.}~\bibnamefont {Savona}}, \bibinfo {author} {\bibfnamefont
  {P.~B.}\ \bibnamefont {Littlewood}}, \bibinfo {author} {\bibfnamefont
  {B.}~\bibnamefont {Deveaud}}, \ and\ \bibinfo {author} {\bibfnamefont
  {L.~S.}\ \bibnamefont {Dang}},\ }\href {\doibase 10.1038/nature05131}
  {\bibfield  {journal} {\bibinfo  {journal} {Nature}\ }\textbf {\bibinfo
  {volume} {443}},\ \bibinfo {pages} {409} (\bibinfo {year}
  {2006})}\BibitemShut {NoStop}%
\bibitem [{\citenamefont {Deng}\ \emph {et~al.}(2010)\citenamefont {Deng},
  \citenamefont {Haug},\ and\ \citenamefont {Yamamoto}}]{Deng2010}%
  \BibitemOpen
  \bibfield  {author} {\bibinfo {author} {\bibfnamefont {H.}~\bibnamefont
  {Deng}}, \bibinfo {author} {\bibfnamefont {H.}~\bibnamefont {Haug}}, \ and\
  \bibinfo {author} {\bibfnamefont {Y.}~\bibnamefont {Yamamoto}},\ }\href
  {\doibase 10.1103/RevModPhys.82.1489} {\bibfield  {journal} {\bibinfo
  {journal} {Reviews of Modern Physics}\ }\textbf {\bibinfo {volume} {82}},\
  \bibinfo {pages} {1489} (\bibinfo {year} {2010})}\BibitemShut {NoStop}%
\bibitem [{\citenamefont {Schneider}\ \emph {et~al.}(2013)\citenamefont
  {Schneider}, \citenamefont {Rahimi-Iman}, \citenamefont {Kim}, \citenamefont
  {Fischer}, \citenamefont {Savenko}, \citenamefont {Amthor}, \citenamefont
  {Lermer}, \citenamefont {Wolf}, \citenamefont {Worschech}, \citenamefont
  {Kulakovskii}, \citenamefont {Shelykh}, \citenamefont {Kamp}, \citenamefont
  {Reitzenstein}, \citenamefont {Forchel}, \citenamefont {Yamamoto},\ and\
  \citenamefont {H{\"{o}}fling}}]{Schneider2013}%
  \BibitemOpen
  \bibfield  {author} {\bibinfo {author} {\bibfnamefont {C.}~\bibnamefont
  {Schneider}}, \bibinfo {author} {\bibfnamefont {A.}~\bibnamefont
  {Rahimi-Iman}}, \bibinfo {author} {\bibfnamefont {N.~Y.}\ \bibnamefont
  {Kim}}, \bibinfo {author} {\bibfnamefont {J.}~\bibnamefont {Fischer}},
  \bibinfo {author} {\bibfnamefont {I.~G.}\ \bibnamefont {Savenko}}, \bibinfo
  {author} {\bibfnamefont {M.}~\bibnamefont {Amthor}}, \bibinfo {author}
  {\bibfnamefont {M.}~\bibnamefont {Lermer}}, \bibinfo {author} {\bibfnamefont
  {A.}~\bibnamefont {Wolf}}, \bibinfo {author} {\bibfnamefont {L.}~\bibnamefont
  {Worschech}}, \bibinfo {author} {\bibfnamefont {V.~D.}\ \bibnamefont
  {Kulakovskii}}, \bibinfo {author} {\bibfnamefont {I.~A.}\ \bibnamefont
  {Shelykh}}, \bibinfo {author} {\bibfnamefont {M.}~\bibnamefont {Kamp}},
  \bibinfo {author} {\bibfnamefont {S.}~\bibnamefont {Reitzenstein}}, \bibinfo
  {author} {\bibfnamefont {A.}~\bibnamefont {Forchel}}, \bibinfo {author}
  {\bibfnamefont {Y.}~\bibnamefont {Yamamoto}}, \ and\ \bibinfo {author}
  {\bibfnamefont {S.}~\bibnamefont {H{\"{o}}fling}},\ }\href {\doibase
  10.1038/nature12036} {\bibfield  {journal} {\bibinfo  {journal} {Nature}\
  }\textbf {\bibinfo {volume} {497}},\ \bibinfo {pages} {348} (\bibinfo {year}
  {2013})}\BibitemShut {NoStop}%
\bibitem [{\citenamefont {Hou}\ \emph {et~al.}(2021)\citenamefont {Hou},
  \citenamefont {Zhang},\ and\ \citenamefont {Liu}}]{Hou2021}%
  \BibitemOpen
  \bibfield  {author} {\bibinfo {author} {\bibfnamefont {J.~T.}\ \bibnamefont
  {Hou}}, \bibinfo {author} {\bibfnamefont {P.}~\bibnamefont {Zhang}}, \ and\
  \bibinfo {author} {\bibfnamefont {L.}~\bibnamefont {Liu}},\ }\href {\doibase
  10.1103/PhysRevApplied.16.034034} {\bibfield  {journal} {\bibinfo  {journal}
  {Physical Review Applied}\ }\textbf {\bibinfo {volume} {16}},\ \bibinfo
  {pages} {034034} (\bibinfo {year} {2021})}\BibitemShut {NoStop}%
\bibitem [{\citenamefont {van~der Pol}(1920)}]{VdP1920}%
  \BibitemOpen
  \bibfield  {author} {\bibinfo {author} {\bibfnamefont {B.}~\bibnamefont
  {van~der Pol}},\ }\href@noop {} {\bibfield  {journal} {\bibinfo  {journal}
  {Philos. Mag.}\ }\textbf {\bibinfo {volume} {2}},\ \bibinfo {pages} {978}
  (\bibinfo {year} {1926})}\BibitemShut {NoStop}%
\bibitem [{\citenamefont {Lee}\ and\ \citenamefont
  {Sadeghpour}(2013)}]{Lee2013}%
  \BibitemOpen
  \bibfield  {author} {\bibinfo {author} {\bibfnamefont {T.~E.}\ \bibnamefont
  {Lee}}\ and\ \bibinfo {author} {\bibfnamefont {H.~R.}\ \bibnamefont
  {Sadeghpour}},\ }\href {\doibase 10.1103/PhysRevLett.111.234101} {\bibfield
  {journal} {\bibinfo  {journal} {Physical Review Letters}\ }\textbf {\bibinfo
  {volume} {111}},\ \bibinfo {pages} {234101} (\bibinfo {year}
  {2013})}\BibitemShut {NoStop}%
\bibitem [{\citenamefont {Dutta}\ and\ \citenamefont
  {Cooper}(2019)}]{Dutta2019}%
  \BibitemOpen
  \bibfield  {author} {\bibinfo {author} {\bibfnamefont {S.}~\bibnamefont
  {Dutta}}\ and\ \bibinfo {author} {\bibfnamefont {N.~R.}\ \bibnamefont
  {Cooper}},\ }\href {\doibase 10.1103/PhysRevLett.123.250401} {\bibfield
  {journal} {\bibinfo  {journal} {Physical Review Letters}\ }\textbf {\bibinfo
  {volume} {123}},\ \bibinfo {pages} {250401} (\bibinfo {year}
  {2019})}\BibitemShut {NoStop}%
\bibitem [{sup()}]{supp}%
  \BibitemOpen
  \href@noop {} {}\bibinfo {howpublished} {See Supplemental Material at
  \url{http://supp} for detailed}\BibitemShut {NoStop}%
\bibitem [{\citenamefont {Klselev}\ \emph {et~al.}(2003)\citenamefont
  {Klselev}, \citenamefont {Sankey}, \citenamefont {Krivorotov}, \citenamefont
  {Emley}, \citenamefont {Schoelkopf}, \citenamefont {Buhrman},\ and\
  \citenamefont {Ralph}}]{Klselev2003}%
  \BibitemOpen
  \bibfield  {author} {\bibinfo {author} {\bibfnamefont {S.~I.}\ \bibnamefont
  {Klselev}}, \bibinfo {author} {\bibfnamefont {J.~C.}\ \bibnamefont {Sankey}},
  \bibinfo {author} {\bibfnamefont {I.~N.}\ \bibnamefont {Krivorotov}},
  \bibinfo {author} {\bibfnamefont {N.~C.}\ \bibnamefont {Emley}}, \bibinfo
  {author} {\bibfnamefont {R.~J.}\ \bibnamefont {Schoelkopf}}, \bibinfo
  {author} {\bibfnamefont {R.~A.}\ \bibnamefont {Buhrman}}, \ and\ \bibinfo
  {author} {\bibfnamefont {D.~C.}\ \bibnamefont {Ralph}},\ }\href {\doibase
  10.1038/nature01967} {\bibfield  {journal} {\bibinfo  {journal} {Nature}\
  }\textbf {\bibinfo {volume} {425}},\ \bibinfo {pages} {380} (\bibinfo {year}
  {2003})}\BibitemShut {NoStop}%
\bibitem [{\citenamefont {Kaka}\ \emph {et~al.}(2005)\citenamefont {Kaka},
  \citenamefont {Pufall}, \citenamefont {Rippard}, \citenamefont {Silva},
  \citenamefont {Russek},\ and\ \citenamefont {Katine}}]{Kaka2005}%
  \BibitemOpen
  \bibfield  {author} {\bibinfo {author} {\bibfnamefont {S.}~\bibnamefont
  {Kaka}}, \bibinfo {author} {\bibfnamefont {M.~R.}\ \bibnamefont {Pufall}},
  \bibinfo {author} {\bibfnamefont {W.~H.}\ \bibnamefont {Rippard}}, \bibinfo
  {author} {\bibfnamefont {T.~J.}\ \bibnamefont {Silva}}, \bibinfo {author}
  {\bibfnamefont {S.~E.}\ \bibnamefont {Russek}}, \ and\ \bibinfo {author}
  {\bibfnamefont {J.~A.}\ \bibnamefont {Katine}},\ }\href {\doibase
  10.1038/nature04035} {\bibfield  {journal} {\bibinfo  {journal} {Nature}\
  }\textbf {\bibinfo {volume} {437}},\ \bibinfo {pages} {389} (\bibinfo {year}
  {2005})}\BibitemShut {NoStop}%
\bibitem [{\citenamefont {Deac}\ \emph {et~al.}(2008)\citenamefont {Deac},
  \citenamefont {Fukushima}, \citenamefont {Kubota}, \citenamefont {Maehara},
  \citenamefont {Suzuki}, \citenamefont {Yuasa}, \citenamefont {Nagamine},
  \citenamefont {Tsunekawa}, \citenamefont {Djayaprawira},\ and\ \citenamefont
  {Watanabe}}]{Deac2008}%
  \BibitemOpen
  \bibfield  {author} {\bibinfo {author} {\bibfnamefont {A.~M.}\ \bibnamefont
  {Deac}}, \bibinfo {author} {\bibfnamefont {A.}~\bibnamefont {Fukushima}},
  \bibinfo {author} {\bibfnamefont {H.}~\bibnamefont {Kubota}}, \bibinfo
  {author} {\bibfnamefont {H.}~\bibnamefont {Maehara}}, \bibinfo {author}
  {\bibfnamefont {Y.}~\bibnamefont {Suzuki}}, \bibinfo {author} {\bibfnamefont
  {S.}~\bibnamefont {Yuasa}}, \bibinfo {author} {\bibfnamefont
  {Y.}~\bibnamefont {Nagamine}}, \bibinfo {author} {\bibfnamefont
  {K.}~\bibnamefont {Tsunekawa}}, \bibinfo {author} {\bibfnamefont {D.~D.}\
  \bibnamefont {Djayaprawira}}, \ and\ \bibinfo {author} {\bibfnamefont
  {N.}~\bibnamefont {Watanabe}},\ }\href {\doibase 10.1038/nphys1036}
  {\bibfield  {journal} {\bibinfo  {journal} {Nature Physics}\ }\textbf
  {\bibinfo {volume} {4}},\ \bibinfo {pages} {803} (\bibinfo {year}
  {2008})}\BibitemShut {NoStop}%
\bibitem [{\citenamefont {Liu}\ \emph {et~al.}(2015)\citenamefont {Liu},
  \citenamefont {Stehlik}, \citenamefont {Eichler}, \citenamefont {Gullans},
  \citenamefont {Taylor},\ and\ \citenamefont {Petta}}]{Liu2015}%
  \BibitemOpen
  \bibfield  {author} {\bibinfo {author} {\bibfnamefont {Y.-Y.}\ \bibnamefont
  {Liu}}, \bibinfo {author} {\bibfnamefont {J.}~\bibnamefont {Stehlik}},
  \bibinfo {author} {\bibfnamefont {C.}~\bibnamefont {Eichler}}, \bibinfo
  {author} {\bibfnamefont {M.~J.}\ \bibnamefont {Gullans}}, \bibinfo {author}
  {\bibfnamefont {J.~M.}\ \bibnamefont {Taylor}}, \ and\ \bibinfo {author}
  {\bibfnamefont {J.~R.}\ \bibnamefont {Petta}},\ }\href {\doibase
  10.1126/science.aaa2501} {\bibfield  {journal} {\bibinfo  {journal}
  {Science}\ }\textbf {\bibinfo {volume} {347}},\ \bibinfo {pages} {285}
  (\bibinfo {year} {2015})}\BibitemShut {NoStop}%
\bibitem [{\citenamefont {Cassidy}\ \emph {et~al.}(2017)\citenamefont
  {Cassidy}, \citenamefont {Bruno}, \citenamefont {Rubbert}, \citenamefont
  {Irfan}, \citenamefont {Kammhuber}, \citenamefont {Schouten}, \citenamefont
  {Akhmerov},\ and\ \citenamefont {Kouwenhoven}}]{Cassidy2017}%
  \BibitemOpen
  \bibfield  {author} {\bibinfo {author} {\bibfnamefont {M.~C.}\ \bibnamefont
  {Cassidy}}, \bibinfo {author} {\bibfnamefont {A.}~\bibnamefont {Bruno}},
  \bibinfo {author} {\bibfnamefont {S.}~\bibnamefont {Rubbert}}, \bibinfo
  {author} {\bibfnamefont {M.}~\bibnamefont {Irfan}}, \bibinfo {author}
  {\bibfnamefont {J.}~\bibnamefont {Kammhuber}}, \bibinfo {author}
  {\bibfnamefont {R.~N.}\ \bibnamefont {Schouten}}, \bibinfo {author}
  {\bibfnamefont {A.~R.}\ \bibnamefont {Akhmerov}}, \ and\ \bibinfo {author}
  {\bibfnamefont {L.~P.}\ \bibnamefont {Kouwenhoven}},\ }\href {\doibase
  10.1126/science.aah6640} {\bibfield  {journal} {\bibinfo  {journal}
  {Science}\ }\textbf {\bibinfo {volume} {355}},\ \bibinfo {pages} {939}
  (\bibinfo {year} {2017})}\BibitemShut {NoStop}%
\bibitem [{\citenamefont {Adler}(1946)}]{Adler1946}%
  \BibitemOpen
  \bibfield  {author} {\bibinfo {author} {\bibfnamefont {R.}~\bibnamefont
  {Adler}},\ }\href {\doibase 10.1109/JRPROC.1946.229930} {\bibfield  {journal}
  {\bibinfo  {journal} {Proceedings of the IRE}\ }\textbf {\bibinfo {volume}
  {34}},\ \bibinfo {pages} {351} (\bibinfo {year} {1946})}\BibitemShut
  {NoStop}%
\end{thebibliography}

\providecommand{\noopsort}[1]{}\providecommand{\singleletter}[1]{#1}%

\end{document}